\documentclass[12pt]{article}
\usepackage{aaspp4}
\usepackage{flushrt}
\usepackage{latexsym}

\newcommand\etal{{et al.\ }}
\newcommand\hb{\mbox{${\rm H}\beta$}}
\newcommand\kms{\mbox{$\ {\rm km\ s}^{-1}$}}
\newcommand\lya{\mbox{${\rm Ly}\alpha$}}
\newcommand\mpc{\mbox{$\ {\rm Mpc}$}}

\begin{document}

\slugcomment{Submitted February 17, 1998; accepted November 19}
\lefthead{\bf CAMPOS ET AL.}
\righthead{\bf A CLUSTER OR FILAMENT OF GALAXIES AT Z=2.5?}

\title{\bf A cluster or filament of galaxies at redshift $\mathbf{z=2.5}$?}

\author{Ana Campos\altaffilmark{1,2}, Amos Yahil\altaffilmark{1,3,4},
Rogier A. Windhorst\altaffilmark{5}, Eric A. Richards\altaffilmark{6}, \\
Sebastian Pascarelle\altaffilmark{3}, Chris Impey\altaffilmark{7}, and
Catherine Petry\altaffilmark{7}}

\altaffiltext{1}{Visiting Astronomer, Kitt Peak National Observatory, National
Optical Astronomy Observatories, which is operated by the Association of
Universities for Research in Astronomy, Inc. (AURA) under cooperative agreement
with the National Science Foundation.}

\altaffiltext{2}{Instituto de Matem\'aticas y F\'{\i}sica Fundamental, CSIC,
Serrano 113bis, E-28006 Madrid, Spain, e-mail: ana@lulu.imaff.csic.es}

\altaffiltext{3}{Department of Physics \& Astronomy, State University of New
York, Stony Brook, NY 11794-3800, USA, e-mail: Amos.Yahil@sunysb.edu (AY),
sam@sbast3.ess.sunysb.edu (SP)}

\altaffiltext{4}{Visiting Scholar, Center for Astrophysics and Space Sciences,
University of California, San Diego, La Jolla, CA 92093-0424, USA}

\altaffiltext{5}{Department of Physics and Astronomy, Arizona State University,
Tempe, AZ 85287-1504, USA, e-mail: raw@cosmos.la.asu.edu.}

\altaffiltext{6}{University of Virginia and National Radio Astronomy
Observatory, USA, e-mail: er4n@virginia.edu}

\altaffiltext{7}{Steward Observatory, University of Arizona, Tucson, AZ 85721,
USA, cimpey@as.arizona.edu (CI), petry@as.arizona.edu (CP).}

\begin{abstract}

  We report the discovery of 56 new \lya-emitting candidates (LECs) at redshift
$z \approx 2.5$ in a field of $8\arcmin\times 14\arcmin$ around two previously
known weak radio QSOs and a cosmic microwave background decrement (CMBD) that
is plausibly due to the Sunyaev-Zel'dovich effect.  Broad-band and medium-band
imaging at the redshifted \lya\ wavelength have allowed us to identify the
LECs at the redshift of the QSOs.  Three of the brightest LECs have been
confirmed spectroscopically, with redshifts between $z=2.501$ and $z=2.557$;
one of them is another QSO.  Excluding the third QSO, the four
spectroscopically confirmed objects form a 3\arcmin\ filament with a rest-frame
velocity dispersion of 1000\kms\ lying adjacent to the CMBD, and there is a
significant concentration of LECs at the NW end of the filament around the
brightest QSO.  If confirmed, a velocity dispersion $\sim 1000\kms$ on a proper
scale $\sim 1\mpc$ at redshift $z=2.5$ would, in and of itself, constrain the
cosmological model to low $\Omega$.

\end{abstract}


\section{INTRODUCTION}

  The abundance of clusters of galaxies is known to evolve at redshifts
$z\lesssim 1$, and this time dependence can be used to constrain cosmological
models \markcite{carl97,bahc97,bart98a} (e.g., Carlberg \etal 1997; Bahcall,
Fan, \& Cen 1997; Bartelmann \etal 1998).  But the nature of clustering at
redshifts $z\gtrsim 1$, where the contrast between model predictions is much
more pronounced, is largely unknown.  While QSO absorption systems show
clustering \markcite{heis89,lanz95,fern96} (e.g., Heisler, Hogan \& White 1989;
Lanzetta, Webb, \& Barcons 1995; Fern\'andez-Soto \etal 1996), and while there
is recent indication of large-scale structure at high redshift
\markcite{stei98,adel98;giav98} (Steidel \etal 1998; Adelberger \etal 1998;
Giavalisco \etal 1998), the identification of specific groups and clusters is
rather tentative
\markcite{dres93,giav94,pasc96a,hu96,fran96,lefe96,warr96,malk96,dick97,stan97}
(Dressler \etal 1993; Giavalisco, Steidel \& Szalay 1994; Pascarelle \etal
1996a; Hu \& MacMahon 1996; Francis \etal 1996; LeF\`evre \etal 1996; Warren \&
M{\o}ller 1996; Malkan, Teplitz, \& McLean 1996; Dickinson 1997; Stanford \etal
1997), and to date no rich cluster has been definitively detected at redshift
$z>1.3$.

  One tracer of a possible distant cluster is a cosmic microwave background
decrement (CMBD), which is most plausibly due to inverse Compton scattering by
intracluster plasma \markcite{suny72} (Sunyaev and Zel'dovich 1972).  Two such
CMBDs without identified optical clusters have recently been reported at
1312+4237 \markcite{rich97} (Richards \etal 1997) and 1643+4631
\markcite{jone97} (Jones \etal 1997).  The absence of X-ray flux detection in
deep ROSAT images can be used to set lower limits for the redshifts of
intervening clusters.  \markcite{knei98} Kneissl, Sunyaev, \& White (1998)
found $z\ge 0.7$ and $z\ge 2.8$ at the 95\% confidence level for the redshifts
of 10-keV isothermal objects of any size along the lines of sight to 1312+4237
and 1643+4631, respectively.

  Intriguingly, both CMBDs are in the direction of QSO pairs at redshifts
$z=2.561$ for 1312+4237 \markcite{wind95} (Windhorst \etal 1995) and $z=3.790$
and $z=3.831$ for 1643+4631 \markcite{schn91} (Schneider, Schmidt, \& Gunn
1991).  The differing redshifts for the 1643+4631 QSOs (corresponding to a
rest-frame velocity difference of 2500\kms) and the large separations
(90\arcsec\ and 200\arcsec\ for 1312+4237 and
1643+4631, respectively) argue against gravitational lensing by
intervening clusters, so it is possible that the clusters are at the redshifts
of the QSOs.

  With this in mind, we searched a field of $8\arcmin\times 14\arcmin$ around
1312+4237 for objects whose medium-band imaging shows enhanced emission
centered at 4350{\rm\AA}.  We found 56 such objects (in addition to the two
previously known QSOs) and interpret them as \lya-emitting candidates (LECs) at
redshift $z\approx 2.5$.  We have confirmed the redshifts of three of the
brightest new LECs (one of which is a QSO), with redshifts between $z=2.501$
and $z=2.557$.  A grey-scale image of the 1312+4237 field, with all the LECs
and the CMBD marked, is shown in Figure \ref{image}.  Data for all the objects
with existing spectroscopy are given in Table 1.

\section{OBSERVATIONS}

  Images of 1312+4237 in the $B$ and $I$ bands (Harris set) were obtained in
1997 April with the prime focus camera of the KPNO 4m telescope (FOV $16'\times
16'$, pixel size 0\farcs47).  Five dithered exposures of 15 minutes each were
taken in each band under photometric conditions with an average seeing of
1\farcs2.  In order to select the LECs, the field was also imaged using a
medium-band filter, $\Delta \lambda=130${\AA}, centered at $4350${\AA}.
(Hereafter we use $B_M$ to designate both the filter and the associated
magnitude.)  A total of sixteen dithered 30-minute exposures were obtained,
giving continuum depth comparable to the $B$ images.  All images were reduced
using standard IRAF procedures, with magnitudes measured by the {\em phot\/}
routine in apertures of 6-pixel radius.  In order to reduce photometric color
errors, the $B$ and the $B_M$ magnitudes were both calibrated against the same
exposures of \markcite{land92} Landolt (1992) standards in $B$.  The zero-point
difference between the magnitudes, calibrated separately with the Landolt
standards, showed a perfectly linear relation, with an {\em rms\/} calibration
error of 0.04 mag.  Color errors were determined by adding in quadrature
magnitude errors.

  The SExtractor program \markcite{bert96} (Bertin \& Arnouts 1996) was used
for automatic source identification; source confusion is not a problem at these
magnitudes.  The presence of a bright star at the east of the field forced the
removal of around 30\% of the image.  A total of 3772 objects with $B \le 26$
were detected in the remaining field of about $8\arcmin \times 14\arcmin$.  We
determined the galaxy catalogue to be complete to $B=25$ and $I=23.5$ by
comparing with published galaxy counts \markcite{metc96} (e.g., Metcalfe \etal
1996).

  The color-magnitude diagram, $B-B_M$ (IRAF aperture photometry) versus
$B_{tot}$ (SExtractor ``best''-magnitude), is shown in Figure \ref{CM} for all
the galaxies detected in the image with magnitudes $B\le 26$ and photometric
color errors $\Delta(B-B_M) < 0.2$ mag.  The most noticeable feature of the
diagram is the number of galaxies with excess color $B-B_M$.

  The traditional method of identifying emission-line candidates has been to
seek objects with color excesses exceeding $2\sigma$ or $3\sigma$, where the
standard deviation includes both the intrinsic spread and the observational
errors \markcite{pasc96b,cowi98} (Pascarelle \etal 1996b; Cowie \& Hu 1998).
But the distribution of observed colors is not Gaussian: {\em intrinsic\/}
colors are skewed toward negative $B-B_M$, as can be seen in Figure \ref{CM}
for galaxies with $B < 23$, for which the sample is complete and observational
errors are negligible, and incompleteness at faint magnitudes affects negative
$B-B_M$ first, since the sample was selected in $B$.  In order to avoid a
complicated selection criterion, we apply a fixed color cut and designate as
LECs all objects with $B-B_M>0.6$ and color errors $\Delta(B-B_M) < 0.2$.  (The
median color errors, after eliminating those exceeding 0.2, are 0.11 and 0.16
mag for $24\le B< 25$ and $25\le B< 26$, respectively.)  To verify the
effectiveness of this cut, we used a variety of evolutionary models
\markcite{camp97} (Campos 1997), convolved with the actual photometric color
errors, to show that, without excess line emission, the number of galaxies
expected to have color excess $B-B_M>0.6$ is very small compared with the
number observed.  As an example, the expectation in an $\Omega_0=0.1$
$\Lambda=0$ model is for less than one galaxy in the magnitude range $24\le B<
25$ to satisfy the color cutoff, $B-B_M>0.6$, and for only 3 galaxies to do so
for $25\le B<26$, while we found 16 and 39 objects, respectively, with colors
exceeding that limit.  Our color criterion, $B-B_M>0.6$, corresponds to an
observed equivalent width over a flat spectrum exceeding 120{\AA}.  At the
$B=25$ completeness limit of our catalogue this translates to a line flux of
$6\times 10^{-17}$ erg cm$^{-2}$ s$^{-1}$ and $B_M=24.4$.  There are 37 LECs
above this limit, or a surface density of $0.33/\Box\arcmin$.  This LEC surface
density is similar to that found by \markcite{stei96} Steidel \etal (1996) for
normal star-forming galaxies at redshifts $z>3$.

  There are 3 objects in the field with magnitudes $B<21.5$ and a clear excess
of emission in the $B_M$ band (see Figure \ref{CM}).  Two of the objects, Q1
and Q2, are the previously known $\mu$Jy QSOs \markcite{wind95} (Windhorst
\etal 1995).  The third object, with similar $B-B_M$ and intermediate $B$
magnitude, was observed at the MMT in 1998 February with a dispersion of
1.95{\AA}/pixel and nominal resolution of 7.1{\AA} and found to be another QSO
at redshift $z=2.501$.  Its spectrum is shown in Figure \ref{Q3}.

  The other LECs are all fainter, $B>24$.  We tried to confirm their redshifts
in another spectroscopic follow-up in 1998 March, using the LDSS-2 multi-slit
spectrograph at the WHT 4m telescope at La Palma Observatory with a dispersion
of 5.3{\AA}/pixel and nominal resolution of 13.3{\AA}.  Inclement weather
allowed us to obtain workable spectra for only 2 objects, G1 and G2.  Both of
them show lines in their spectra that, if identified with \lya, imply redshifts
of $z=2.557$ and $z=2.536$, respectively.  Their spectra are plotted in Figure
\ref{GALS}.  Note that both galaxies show little or no emission shortward of
the putative \lya\ line and continuum longward of the line, consistent with a
\lya-forest interpretation.  With our signal-to-noise ratio and resolution we
cannot exclude the possibility that they are also active galaxies.  In three
other objects with $0.60 < B-B_M < 0.65$ we could barely detect continua but no
lines.  One object close to G1 in the color-magnitude diagram, Figure \ref{CM},
but with $B-I=1.9$, was not detected at all.  The spectroscopic confirmation
rate is similar to the one in the 53W002 field of \markcite{pasc96b} Pascarelle
\etal (1996b).

\section{DISCUSSION}

  In the 1312+4237 field under analysis---$8\arcmin\times 14\arcmin$ around two
known QSOs at redshift $z=2.561$ and a cosmic microwave background decrement
(CMBD)---we have found 56 new \lya-emitting candidates (LECs) at redshift
$z\approx 2.5$.  Three of the new LECs now have confirmed redshifts ranging
from $z=2.501$ to $z=2.557$; one of them is another QSO at $z=2.5$.

  It is unlikely that the excess emission in the as yet unconfirmed LECs is due
to another line, for example [OII]$\lambda 3727$ at redshift $z=0.16$.  First,
\hb\ was not observed at the expected wavelengths in the spectra of G1 and G2
(dotted lines in Figure \ref{GALS}).  Second, at redshift $z=0.16$ the galaxies
would be starbursting dwarfs with $M_B\sim -15$, W[OII]$\gtrsim 80${\AA}, and
density $\sim 6\times 10^{-2}\ h_{50}^3 \mpc^{-3}$.  For any cosmology, the
space density of such a population implied by our observation would far exceed
that observed heretofore \markcite{camp97} (e.g., Campos 1997).  Moreover, many
of the LECs are red (31 with colors $B-I\ge 1$), while starbursting dwarfs are
expected to be blue, $0 \la B-I \la 0.5$.

  The LECs are scattered throughout the observed field of $8\arcmin\times
14\arcmin$, corresponding to proper lengths of $6\times 10\ h_{50}^{-1} \mpc$
for $\Omega_0=0.1$ and $\Lambda=0$ and a factor of 1.5 smaller if $\Omega_0=1$.
Virialized clusters at the present epoch are somewhat smaller, but some of the
LECs may also belong to a surrounding supercluster and not be part of a
virialized cluster.  For non-virialized objects, such as protoclusters and
filaments, the relevant scales can be as high as the comoving dimensions of the
field, $20\times 35\ h_{50}^{-1}\mpc$, corresponding to present-day large-scale
structure.  Using the same medium-band imaging technique at 4100\AA,
\markcite{pasc98} Pascarelle, Windhorst, \& Keel (1998) find densities ranging
from $0.2-2/\Box\arcmin$ in four fields complete to a depth of 25 mag, three of
which were randomly chosen.  Multiplying our surface density of
$0.33/\Box\arcmin$ by a factor of 1.4 to bring it to the same completeness
level, we find it to be in mid-range.  The galaxian density may not reflect the
underlying mass density, however.  Reported biasing factors at high redshifts
range from high \markcite{adel98} (Adelberger \etal 1998) to extreme
\markcite{mann98} (Mannucci \etal 1998).

  Four of the five spectroscopically confirmed objects actually form a
3\arcmin\ filament with a rest-frame velocity dispersion of 1000\kms\ located
near the CMBD, and the CMBD possibly extends NE across it \markcite{wind95}
(Windhorst \etal 1995, Figure 1).  (Q3 is 10\arcmin\ away and has a relative
velocity of 4500\kms.)  Moreover, five LECs with $0.75 \pm 0.11 < B-B_M < 1.63
\pm 0.12$, including G2, are within 0.7\arcmin\ of Q1 at the NW end of the
filament.  Since only half the LECs have $B-B_M > 0.75$, their mean surface
density in the field is $0.25/\Box\arcmin$, and the probability for such a
clustering to be random is $5\times 10^{-5}$.  This probability need not be
viewed as completely {\em a posteriori}, since Q1 is by far the brightest LEC
in the field and may well point to the location of the cluster (but note that
some of the LECs may be due to the QSO itself, cf., \markcite{nata98}
Natarajan, Sigurdsson, \& Silk 1998).

  Deeper imaging might reveal more structure, but spectroscopy is the critical
confirmatory observation.  While cluster evolution is usually stated in terms
of the abundances of properly identified clusters, velocity dispersion is a
dynamical measure of mass in its own right and can be compared directly with
numerical simulations.  A velocity dispersion $\sim 1000\kms$ on a proper scale
$\sim 1\mpc$ at redshift $z=2.5$ would, in and of itself, constrain
cosmological models to low $\Omega$ \markcite{eke96} (Eke, Cole, \& Frenk
1996), as would a definitive association of the CMBD with this redshift
\markcite{bart98b} (Bartlett, Blanchard, \& Barbosa 1998).  If confirmed, the
1312+4237 cluster would therefore be a ``smoking-gun'' measure of $\Omega$.

\acknowledgements

  We thank M. Moles for a useful discussion and a careful reading of the
manuscript.  Part of this work was supported by NASA grants AR-07551.01-96A (to
AY), and GO-5985.01-94A, GO-6610.01-95A, and GO.2684.03-94A (to RAW) from
STScI, which is operated by AURA, Inc., under NASA contract NAS5-26555.

\newpage

\section*{FIGURE CAPTIONS}

\figcaption{Grey-scale $15\arcmin \times 14\arcmin$ image of 1312+4237 in the
medium band, $B_M$, with north and east to the top and left as usual.  Circles
mark the position of the two previously known QSOs, Q1 and Q2, and the new one,
Q3.  Squares and triangles mark the other LECs with $B-I$ color $\geq 1$ and
$<1$, respectively, with the two spectroscopically confirmed galaxies
identified as G1 and G2.  The ellipse is at the position of the cosmic
microwave background decrement.\label{image}}

\figcaption{$B-B_M$ versus $B$ color-magnitude diagram for all the galaxies
detected with magnitude $B\le 26$ and color photometric error
$\Delta(B-B_M)<0.2$.  The objects marked as Q1 and Q2 are the two known $\mu$Jy
QSOs.  Q3, G1 and G2 denote the newly confirmed QSO and two
galaxies.\label{CM}}

\figcaption{Spectrum of the newly confirmed QSO, Q3.\label{Q3}}

\figcaption{Spectrum of the newly confirmed galaxies, G1 and G2.  Note that if
the emission lines are O[II]$\lambda 3727$ at redshift $z=0.16$, then \hb\
emission is expected at the positions of the dotted lines but is not
seen.\label{GALS}}

\newpage

\begin{deluxetable}{ccccccc}
\tablenum{1}
\tablewidth{0pt}
\tablecaption{Spectroscopically Confirmed \lya-Emitting Objects}
\tablehead{
\colhead{Object}		& \colhead{$\alpha$ (J2000)}	&
\colhead{$\delta$ (J2000)}	& \colhead{$B$}			&
\colhead{$B-B_M$}		& \colhead{$B-I$}		&
\colhead{$z$}
}
\startdata
Q1 & 13 12 15.3 & 42 39 00 & 18.57 & 0.75$\pm$0.01 & 0.66$\pm$0.01 & 2.561 \nl
Q2 & 13 12 22.4 & 42 38 14 & 21.19 & 0.69$\pm$0.01 & 1.57$\pm$0.01 & 2.561 \nl
Q3 & 13 12 39.4 & 42 29 40 & 20.54 & 0.79$\pm$0.01 & 0.73$\pm$0.01 & 2.501 \nl
G1 & 13 12 25.1 & 42 37 23 & 24.38 & 0.75$\pm$0.09 & 0.28$\pm$0.33 & 2.557 \nl
G2 & 13 12 11.6 & 42 39 06 & 24.67 & 1.15$\pm$0.10 & 1.00$\pm$0.21 & 2.536 \nl
\enddata
\end{deluxetable}


\newpage

\begin{references}

\reference{adel98} Adelberger, K. L., Steidel, C. C., Giavalisco, M.,
Dickinson, M. Pettini, M., \& Kellogg, M. 1998, \apj, 505, 18

\reference{bahc97} Bahcall, N. A., Fan, X., Cen, R. 1997, \apjl, 485, L53

\reference{bart98a} Bartelman, M., Huss, A., Colberg, J. M., Jenkins, A., \&
Pearce, F. R. 1998, \aap, 330, 1

\reference{bart98b} Bartlett, J. G., Blanchard, A., \& Barbosa, D. 1998, \aap,
336, 425

\reference{bert96} Bertin, E., \& Arnouts, S. 1996 \aap, 117, 393

\reference{camp97} Campos, A. 1997, \apj, 488, 606

\reference{carl97} Carlberg, R. G., Morris, S. L., Yee, H. K. C., \& Ellingson,
E. 1997, \apjl, 479, L19

\reference{cowi98} Cowie, L. L., \& Hu, E. M. 1998, \aj, 115, 1319

\reference{dick97} Dickinson, M. 1997, in ``The Hubble Space Telescope and the
High Redshift Universe'', ed.\ N. R. Tanvir, A. Arag\'on-Salamanca, \&
J. V. Wall, (Singapore: World Scientific), 207

\reference{dres93} Dressler, A., Oemler Jr., A., Gunn, J. E., \& Butcher,
H. 1993, \apjl, 404, L45

\reference{eke96} Eke, V. R., Cole, S., \& Frenk, C. S. 1996, \mnras, 282, 263

\reference{fern96} Fern\'andez-Soto, A., Lanzetta, K. M., Barcons, X.,
Carswell, R. F., Webb, J. K., \& Yahil, A. 1996, \apjl, 460, L85

\reference{fran94} Francis, P. J., \etal 1994, \apj, 457, 490

\reference{giav94} Giavalisco, M. Steidel, C. C., \& Szalay, A. S. 1994, \apjl,
425, L5

\reference{giav98} Giavalisco, M., Steidel, C. C., Adelberger, K. L.,
Dickinson, M., Pettini, M., \& Kellogg, M. 1998, \apj, 503, 543

\reference{heis89} Heisler, J., Hogan, C.J. \& White, S.D.M. 1989, \apj, 347,
52.

\reference{hu96} Hu, E. M., \& MacMahon, R. G. 1996, \nat, 382, 231

\reference{jone97} Jones, M. E., \etal 1997, \apjl, 479, L1

\reference{knei98} Kneissl, R., Sunyaev, R. A., \& White, S. D. M. 1998,
\mnras, 297, L29

\reference{land92} Landolt, A. U. 1992, \aj, 104, 340

\reference{lanz96} Lanzetta, K. M., Webb, J. K., \& Barcons, X., 1996, \apjl,
456, 17

\reference{lefe96} LeF\`evre, O., Deltorn, J.-M., Crampton, D., \& Dickinson,
M. 1996, \apjl, 471, L11

\reference{malk96} Malkan, M. A., Teplitz, H., \& McLean, I. S. 1996, \apjl, 468, L9

\reference{mann98} Mannucci, F., Thompson, D., Beckwith, S. V. W., and Williger, G. M. 1998, \apjl, 501, L11.

\reference{metc96} Metcalfe, N., Shanks, T., Campos, A., Fong, R., \& Gardner,
J. 1996, \nat, 383, 236

\reference{nata98} Natarajan, P., Sigurdsson, S., \& Silk, J. 1998, \mnras,
298, 577

\reference{pasc96a} Pascarelle, S. M., Windhorst, R. A., Driver, S. P., \&
Ostrander, E. J. 1996a, \apjl, 456, L21

\reference{pasc98} Pascarelle, S. M., Windhorst, R. A., \& Keel, W. C. 1998,
\apj, in press; preprint astro-ph/09181
 
\reference{pasc96b} Pascarelle, S. M., Windhorst, R. A., Keel, W. C., \&
Odewahn, S. C. 1996b, \nat, 383, 45

\reference{rich97} Richards, E. A., Fomalont, E. B., Kellerman, K. I.,
Partridge, R. B., \& Windhorst, R. A. 1997, \aj, 113, 1475

\reference{schn91} Schneider, D. P., Schmidt, M., \& Gunn, J. E. 1991, \aj,
101, 2004

\reference{stan97} Stanford, S. A., Elston, R., Eisenhardt, P. R., Spinrad, H.,
Stern, D., \& Dey, A. 1997, AJ, 114, 2232

\reference{stei98} Steidel, C. C., Adelberger, K. L., Dickinson, M.,
Giavalisco, M., Pettini, M. \& Kellogg, M. 1998, \apj, 492, 428.

\reference{stei96} Steidel, C. C., Giavalisco, M., Pettini, M., Dickinson, M.,
\& Adelberger, K. L. 1996, \apjl, 462, L17

\reference{suny72} Sunyaev, R. A., \& Zel'dovich, Ya. 1972, Comments
Astrophys. Space Phys., 4, 173

\reference{warr96} Warren, S. J., \& M{\o}ller, P. 1996, \aap, 311, 25

\reference{wind95} Windhorst, R. A., Fomalont, E. B., Kellerman, K. I.,
Partridge, R. B., Richards, E., Franklin, B. E., Pascarelle, S. M., \&
Griffiths, R. E. 1995, \nat, 375, 471

\end{references}
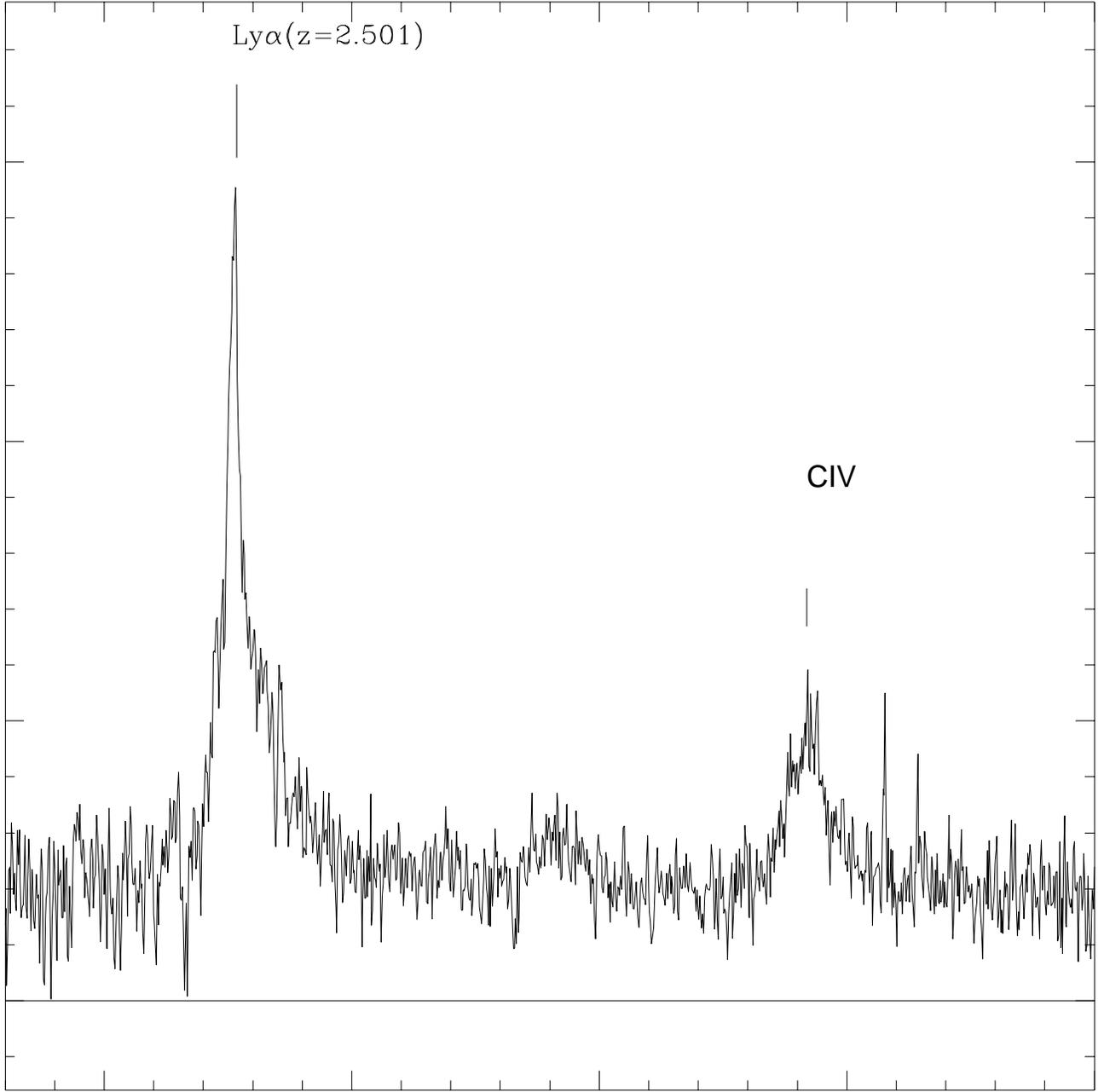
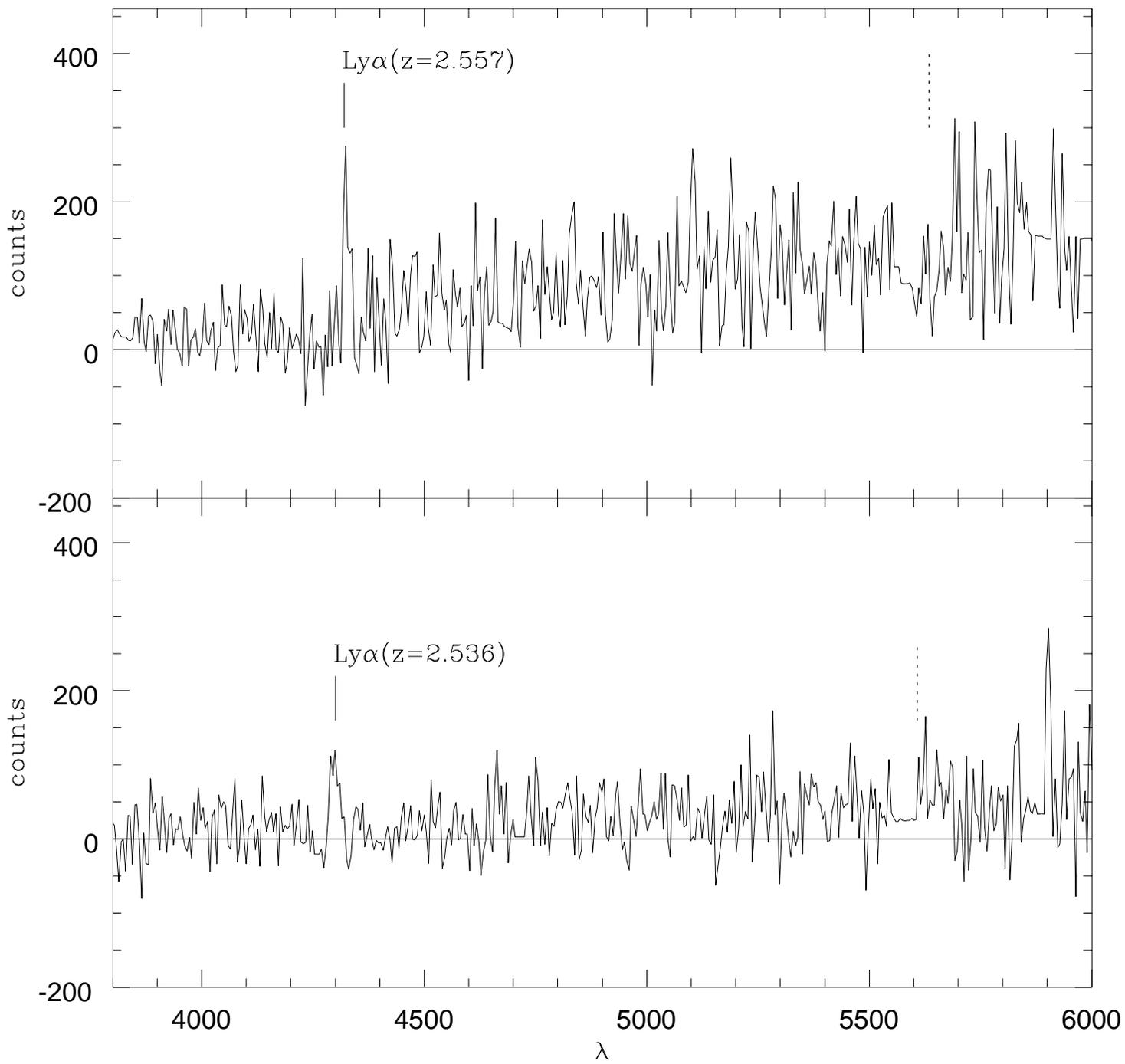
\end{document}